\documentclass{article}

\usepackage{PRIMEarxiv}

\usepackage[utf8]{inputenc} 
\usepackage[T1]{fontenc}    
\usepackage{hyperref}       
\usepackage{url}            
\usepackage{booktabs}       
\usepackage{amsfonts}       
\usepackage{nicefrac}       
\usepackage{microtype}      
\usepackage{lipsum}
\usepackage{fancyhdr}       
\usepackage{graphicx}       
\graphicspath{{media/}}     

\usepackage{todonotes}

\usepackage{soul}

\usepackage[natbib, bibencoding=utf8, citestyle=numeric, bibstyle=ieee, maxbibnames=999, maxcitenames=2, mincitenames=1, sortcites]{biblatex}
\bibliography{references.bib}

\usepackage[acronym, shortcuts, nohypertypes={acronym}]{glossaries}

\newacronym{AI}{AI}{artificial intelligence}
\newacronym{DL}{DL}{deep learning}
\newacronym{DNN}{DNN}{deep neural network}
\newacronym{GAN}{GAN}{generative adversarial network}
\newacronym{ML}{ML}{machine learning}
\newacronym{NLP}{NLP}{natural language processing}

\usepackage[capitalise, nameinlink, noabbrev]{cleveref}

\title{
HEAR4Health: A blueprint for making computer audition a staple of modern healthcare
}

\author{
\textbf{Andreas Triantafyllopoulos\textsuperscript{1}}, 
\textbf{Alexander Kathan\textsuperscript{1},}
\textbf{Alice Baird\textsuperscript{1}},
\textbf{Lukas Christ\textsuperscript{1}},
\textbf{Alexander Gebhard\textsuperscript{1},} \\
\textbf{Maurice Gerczuk\textsuperscript{1},}
\textbf{Vincent Karas\textsuperscript{1},}
\textbf{Tobias H\"{u}bner\textsuperscript{1},}
\textbf{Xin Jing\textsuperscript{1},}
\textbf{Shuo Liu\textsuperscript{1},}\\
\textbf{Adria Mallol-Ragolta\textsuperscript{1,3},}
\textbf{Manuel Milling\textsuperscript{1},}
\textbf{Sandra Ottl\textsuperscript{1},}
\textbf{Anastasia Semertzidou\textsuperscript{1},}\\
\textbf{Srividya Tirunellai Rajamani\textsuperscript{1},}
\textbf{Tianhao Yan\textsuperscript{1},}
\textbf{Zijiang Yang\textsuperscript{1},} 
\textbf{Judith Dineley\textsuperscript{1},} \\ 
\textbf{Shahin Amiriparian\textsuperscript{1},}
\textbf{Katrin D. Bartl-Pokorny\textsuperscript{1,2},}
\textbf{Anton Batliner\textsuperscript{1},} 
\textbf{Florian B. Pokorny\textsuperscript{1,2,3},}
 \textbf{Bj\"{o}rn W. Schuller\textsuperscript{1,3,4}}\\\\
\textsuperscript{1} EIHW -- Chair of Embedded Intelligence for Healthcare and Wellbeing, University of Augsburg, Germany\\
\textsuperscript{2} Division of Phoniatrics, Medical University of Graz, Austria\\
\textsuperscript{3} Centre for Interdisciplinary Health Research, University of Augsburg, Germany\\  
\textsuperscript{4} GLAM -- Group on Language, Audio, \& Music, Imperial College London, United Kingdom\\
}

\begin{document}
\maketitle

\begin{abstract}
Recent years have seen a rapid increase in digital medicine research in an attempt to transform traditional healthcare systems to their modern, intelligent, and versatile equivalents that are adequately equipped to tackle contemporary challenges.
These efforts are being spearheaded by advances in artificial intelligence (AI), and specifically in intelligent sensing technologies that assist doctors in the diagnosis and subsequent monitoring of diseases.
This has led to a wave of applications that utilise AI technologies; first and foremost in the fields of medical imaging, but also in the use of wearables and other intelligent sensors.
In comparison, computer audition can be seen to be lagging behind, at least in terms of commercial interest.
Yet, audition has long been a staple assistant for medical practitioners, with the stethoscope being the quintessential sign of doctors around the world.
Transforming this traditional technology with the use of AI entails a set of unique challenges.
We categorise the advances needed in four key pillars: \underline{H}ear, corresponding to the cornerstone technologies needed to analyse auditory signals in real-life conditions; \underline{E}arlier, for the advances needed in computational and data efficiency; \underline{A}ttentively, for accounting to individual differences and handling the longitudinal nature of medical data; and, finally, \underline{R}esponsibly, for ensuring compliance to the ethical standards accorded to the field of medicine.
Naturally, these advances do not exist in a vacuum, but are inextricably tied to the applications we are targeting -- sensing a range of medical conditions whose manifestation in the audio signal makes them conducive to the use of audition.
Thus, we provide an overview and perspective of HEAR4Health: the sketch of a modern, ubiquitous sensing system which can bring computer audition on par with other AI technologies in the strive for improved healthcare systems.
\end{abstract}

\keywords{Computer Audition \and Digital Health \and Digital Medicine}

\section{Introduction}
Following the rapid advancements in \ac{AI}, and in particular those related to \ac{DL}~\citep{LeCun15-DL}, digital health applications making use of those technologies are accordingly on the rise.
Most of them are focused on diagnosis: 
from computer vision techniques applied to digital imaging~\citep{Esteva21-DLE} to wearable devices monitoring a variety of signals~\citep{Amft18-HWC, Tu20-EDH}, \ac{AI} tools are being increasingly used to provide medical practitioners with a more comprehensive view of their patients.
Computer audition complements this assortment of tools by providing access to the audio generated by a patient's body.
Most often, this corresponds to speech produced by the patients -- sometimes natural, mostly prompted~\citep{Cummins18-SAH, latif2020speech, Milling22-SNB}.
However, there exists a plethora of auditory signals emanating from the human body, all of which are potential carriers of information relating to disease.

Acquiring those auditory biosignals is the first, crucial step in a computer audition pipeline.
Oftentimes, this must be done in noisy environments where audio engineers have little to no control, e.\,g., in a busy hospital room or the patient's home.
This results in noisy, uncurated signals which must be pre-processed in order to become usable, a process which is extremely laborious if done manually.
Automating this process becomes the domain of the first of four outlined pillars, \textbf{(I) Hear}, which is responsible for denoising, segmenting, and altogether preparing the data for further processing by the downstream algorithms.

Those algorithms typically comprise learnable components, i.\,e., 
functions whose parameters are going to be learnt from the consumed data; in the current generation of computer audition systems, the backbone of those algorithms consists of \ac{DL} models.
These models, in turn, are typically very data `hungry', and require an enormous amount of computational resources and experimentation to train successfully.
However, in the case of healthcare applications, such data might not exist, either due to privacy regulations which prohibit their open circulation, or, as in the case of rare or novel diseases, simply because this data does not exist.
Yet doctors, and subsequently the tools they use, are commonly required to operate in such low-data regimes.
Therefore, it is imperative to make these algorithms operational \textbf{(II) Earlier} than what is currently possible; this can be done, for example, by transferring knowledge from domains where data is widely available to data-sparse healthcare applications.

The first two pillars are of a more `engineering' nature; the third one requires more theoretical advances.
Statistical learning theory, which forms the foundation of \ac{DL}, is based on the core assumption that data are \textit{independent and identically distributed}~\citep{shalev2014understanding}.
In the healthcare domain, this translates that the population of training patients is representative of the entire population -- an assumption that often does not hold in practice.
Instead, patients come from different backgrounds, and are typically organised in sub-populations.
Oftentimes, the level of analysis reaches all the way down to the individual; in this case, every patient is considered `unique'.
Furthermore, the larger upside of using \ac{AI} in medicine lies in providing more fine-grained information in the form of longitudinal observations.
Handling the need for individualised analysis with multiple observations over time requires algorithms to operate \textbf{(III) Attentively} to individual -- often changing -- needs.

The last pillar corresponds to the translation of the mandate enshrined in the Hippocratic oath to computer audition, and more generally \ac{AI}: any developed technologies must be developed and be able to operate \textbf{(IV) Responsibly}.
The responsibility lies with the developers and users of the technology and is targeted towards the patients who become its objects.
This informs a set of guidelines and their accompanying technological innovations on how data needs to be sourced, how algorithms must meet certain fairness requirements, and, ultimately, on `doing good' for mankind.

Finally, we would be amiss not to mention the potential applications that can benefit from the introduction of computer audition in healthcare.
This becomes the central component which permeates all aspects of the four pillars: they exist insofar as they serve the overarching goal of providing medical practitioners with novel tools that can help them understand, analyse, diagnose, and monitor their patients' \textbf{Health}.

\begin{figure}[t]
    \centering
    \includegraphics[width=\textwidth]{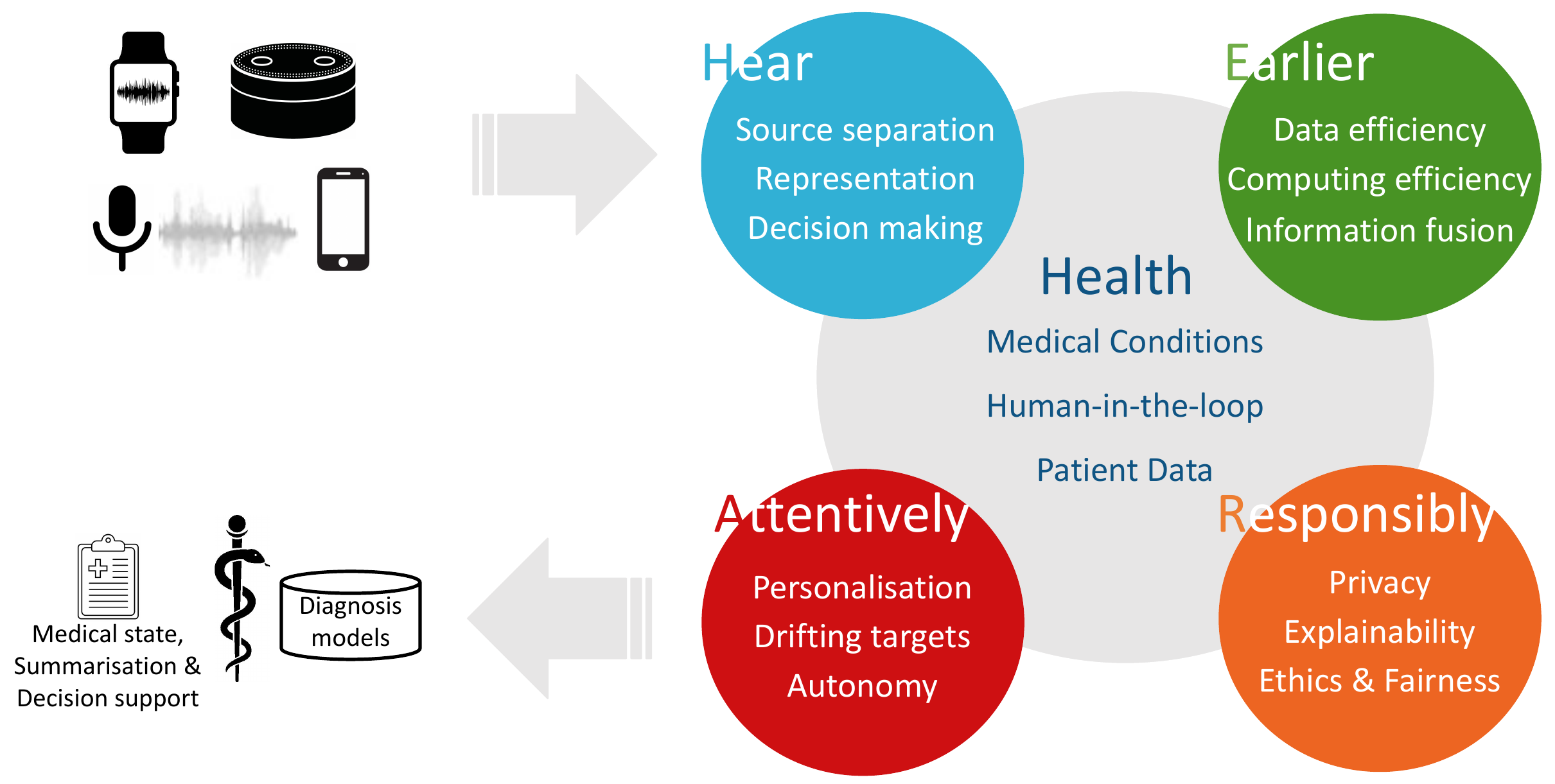}
    \caption{
    Overview of the four pillars for computer audition in healthcare.
    }
    \label{fig:HEAR}
\end{figure}

An overview of the four pillars, as well as their interactions with one another is shown in \cref{fig:HEAR}.
In the following sections, we proceed to analyse each one in more detail, and close with an overview of the healthcare applications in which we expect computer audition to make a decisive difference.
Thus, we present \textbf{HEAR4Health}: a blueprint of what needs to be done for audition to assume its rightful place in the toolkit of \ac{AI} technologies that are rapidly revolutionising healthcare systems around the world.

\section{Hear}

A cornerstone of computer audition applications for healthcare is the ability to \emph{Hear}: that is, the set of steps required to capture and pre-process audio waves and transform them into a clear, useful, and high-quality signal.
This is all the more true in the healthcare domain, where recordings are often made in hospital rooms bustling with activity or conducted at home by the non-expert users themselves.
Therefore, the first fundamental step in an application is to extract only the necessary components of a waveform.

In general, this falls under a category of problems commonly referred to as \emph{source separation and diarisation}~\citep{anguera2012speaker, wang2018supervised}: the separation part corresponds to the extraction of a signal coming from a particular source amongst a mixture of potentially overlapping sources, whereas diarisation corresponds to the identification of temporal start and end times of components assigned to specific subjects.
In healthcare applications, these target components are the relevant sounds; this can include vocalisations (both verbal and non-verbal) but also other bodily sounds that can be captured by specialised auditory sensors attached to their body, or general ones that are monitoring the environment.
These sounds need to be separated from all other sources; these may include a medical practitioner's own body sounds (e.\,g., their voice in doctor-patient conversations) or background environmental noise (e.\,g., babble noise in a hospital).
Accordingly, successful preparation entails a) the ability to recognise which sounds belong to the target subject, b) the ability to detect their precise start and end times, and c) the ability to remove all other signals that co-occur during that time from the waveform.

Traditionally, these steps are tackled by specialised pipelines, which include learnable components that are optimised in supervised fashion~\citep{wang2018supervised}.
For example, the ability to recognise which sounds belong to the target subject is generally referred to as \emph{speaker identification}~\citep{snyder2018x}.
While this term is usually reserved for applications where speech is the sound of interest, it can also be generalised to other bodily sounds~\citep{jokic2022tripletcough}.

Similarly, separation is typically done in a supervised way~\citep{wang2018supervised}.
During the training phase, clean audio signals are mixed with different noises, and a network is trained to predict the original, clean signal from the noisy mixture.
As generalisability to new types of noise sources is a necessary pre-requisite, researchers often experiment with test-time adaptation methods, which adaptively configure a separation model to a particular source~\citep{liu2021n}.

The crucial role of the \emph{Hear} pillar becomes evident when considering data collection.
There are three main data collection paradigms employed in healthcare applications: 
a) the (semi-)structured doctor-patient interview,
b) ecological momentary assessments (EMAs) based on prompts~\citep{shiffman2008ecological},
and, c) passive, continual monitoring~\citep{cornet2018systematic}.
All of them require very robust patient identification and diarisation capabilities.

\section{Earlier}
The major promise of digital health applications is their ubiquitous presence, allowing for a much more fine-grained monitoring of patients than was possible in the past.
This requires the systems to work on mobile devices in an energy-efficient way.
Additionally, these systems must be versatile, and easy to update in the case of new diseases, such as COVID-19.
This requires them to generalise well while being trained on very scarce data.
However, training state-of-the-art \ac{DL} models is a non-trivial process, in many cases requiring weeks or even months, and is furthermore notoriously data intensive. 
Moreover, the technology required, such as high-end GPUs, is often expensive and has exceptionally high energy consumption~\citep{strubell2019energy}.

There have consequently been increasing efforts to develop AutoML approaches that optimise a large network until it is executable on a low-resource device~\citep{cheng2017survey,Amiriparian22-DAP}. 
Many of these approaches focus on reducing the memory footprint and the computational complexity of a network while preserving its accuracy. 
These techniques have shown promise across a range of different learning tasks, however, their potential has not yet been realised for audio-based digital health applications.

On the issue of data efficiency, there has been a lot of research on utilising transfer learning techniques for increasing performance and decreasing the required amount of data.
This is usually done by transferring knowledge from other tasks~\citep{guedes2019transfer, sertolli2021representation}, or even other modalities~\citep{Amiriparian17-SSC,Amiriparian19-DRL}.
However, in the case of audio in particular, an extra challenge is presented by the mismatch between the pre-training and downstream domains~\citep{Triantafyllopoulos21-RTA}.
Recently, large models pre-trained in self-supervised fashion have reached exceptional performance on a variety of different downstream tasks, including the modelling of respiratory diseases~\citep{Triantafyllopoulos22-DBP}, while showing more desirable robustness and fairness properties~\citep{Wagner22-DOT}.

The implementation details of the \emph{Earlier} pillar largely depend on the biomarkers related to the specific medical condition of interest.
For example, in terms of mental disorders, which mostly manifest as pathologies of speech and language, it is mostly tied to generalisation across different languages.
On the one hand, linguistic content itself is a crucial biomarker; on the other hand, it serves to constrain the function of acoustic features; thus, there is a need to learn multi-lingual representations that translate well to low-resource languages.
For diseases manifesting in sounds other than speech signals, the \emph{Earlier} pillar would then improve the data efficiency of their categorisation.
For example, contrary to speech signals, for which large, pre-trained models are readily available~\citep{baevski2020wav2vec}, there is a lack of similar models trained on cough data; a lack partially attributable to the dearth of available data.
This can be overcome, on the one hand, through the use of semi-supervised methods that crawl data from public sources~\citep{amiriparian2017cast}, and, on the other hand, by pursuing (deep) representation learning methods tailored to cough sound characteristics.

When COVID-19 took the world by storm in early 2020, it represented a new, previously unseen threat for which no data was available.
However, COVID-19 is `merely' a coronavirus targeting the upper and lower respiratory tracts, thus sharing common characteristics with other diseases in the same family ~\citep{zou2020sarscov2}.
Transferring prior knowledge from those diseases, while rapidly adapting to the individual characteristics of COVID-19, can be a crucial factor when deploying auditory screening tools in the face of a pandemic.

\section{Attentively}
\label{sec:attentively}
Most contemporary digital health applications focus on the identification of subject states in a static setting, where it is assumed that subjects belong to a certain category or have an attribute in a certain range.
However, many conditions have symptoms that manifest gradually~\citep{amieva2008prodromal}, which makes their detection and monitoring over time a key proposition for future digital health applications. 
Furthermore, disease emergence and progression over time can vary between individuals~\citep{wilson2002individual, pinto2020prediction, hizel2017introduction}. 
For example, the age at onset and the progression rate of age-related cognitive decline varies between individuals~\citep{wilson2002individual}, while there is substantial heterogeneity in the manifestation and development of (chronic) cough across different patients~\citep{mazzone2018heterogeneity}. 
Focusing on these aspects of digital health by adapting to changes in distributions and developing personalised approaches can drastically improve performance.

Recent \ac{DNN}-based methods for personalised \ac{ML}~\citep{rudovic2018personalized} and speaker adaptation~\citep{triantafyllopoulos2021deep} already pave the way for creating individualised models for different patients. 
However, these methods are still in their nascent stage in healthcare~\citep{chen2021personalized}.
Personalised \ac{ML} is a paradigm which attempts to jointly learn from data coming from several individuals while accounting for differences between them. 
Advancing this paradigm for speech in digital health by utilising longitudinal data from several patients for learning to track changes in vocal and overall behaviour over time is a necessary precondition for the digital health systems of the future. 
This means that time-dependent, individualised distributions are taken into account for each patient, by that requiring the development of novel techniques better suited to the nature of this problem; in particular, developing versatile \ac{DL} architectures consisting of global components that jointly learn from all subjects, and specialised ones which adapt to particular patients~\citep{Gerczuk2022-PDL, kathan2022personalised}. 
This novel framing will also enable faster adaptation to new patients by introducing and adapting new models for those patients alone.

\section{Responsibly}
The development of responsible digital health technology is a key pillar of future healthcare applications. 
This ensures trustworthiness and encourages the adherence of users to monitoring protocols. 
Consequently, addressing crucial factors and technology-related consequences in automated disease detection concerning human subjects in a real-world context is of paramount importance.

This pillar intersects with all previous ones and informs their design, adhering to an `ethical-by-design' principle which is fundamental for healthcare applications.
Naturally, a first requirement that applies to all pillars is one of evaluation: all components of a healthcare application need to be comprehensively evaluated with respect to all sub-populations and sensitive attributes.
This holds true for all components of a computer audition system: from extracting the target audio signal (Hear) to generating efficient representations (Earlier) and adapting to individual characteristics (Attentively), any developed methods should perform equally for different sub-populations.
The evaluation could be complemented by explainability methods, which explicitly search for biases in model decisions~\citep{arrieta2020explainable}.

Aside from comprehensively evaluating all methods with respect to fairness, explicit steps must be taken to improve on those~\citep{du2020fairness}.
To this end, adversarial~\citep{wang2019balanced} and constraint-based methods~\citep{zafar2017fairness} have been proposed to learn fair representations. 
In adversarial debiasing, the main predictive network learns to perform its task while an adversary pushes it toward representations which obfuscate the protected characteristics.
Constraint-based methods instead solve the main prediction task subject to fairness constraints (such as equality of opportunity); these methods rely on convex relaxation or game-theoretic optimisation to efficiently optimise the constrained loss function.

The second requirement placed on the three other pillars is privacy.
For example, the \emph{Hear} pillar could be co-opted to remove private information (e.\,g., via using keyword spotting to remove sensitive linguistic information).
The \emph{Earlier} pillar would then take the extracted signal and remove any paralinguistic information unrelated to the task; this could be achieved by targeted voice conversion that preserves any required signal characteristics but changes the patient's voice to be unrecognisable~\citep{jordon2018pate}. 

Satisfying this requirement, however, is particularly challenging for the \emph{Attentively} pillar, as there is a natural privacy-personalisation trade-off: the more private information is removed, the less context remains to be utilised for the target patient.
The main solution to this obstacle is the use of federated learning~\citep{fallah2020personalized}: to ensure that sensitive information cannot be derived from central models, differential privacy methods have been proposed, such as differentially private stochastic gradient descent~\citep{song2013stochastic} and a private aggregation of teacher ensembles~\citep{papernot2018scalable}.
These methods would update the global model backbone discussed in \cref{sec:attentively}, which is shared among all `clients', while any personalised components would remain local -- and thus under the protection of safety mechanisms implemented by the client institutions.


\section{Healthcare Applications}
\begin{figure}
    \centering
    \includegraphics[width=0.7\textwidth]{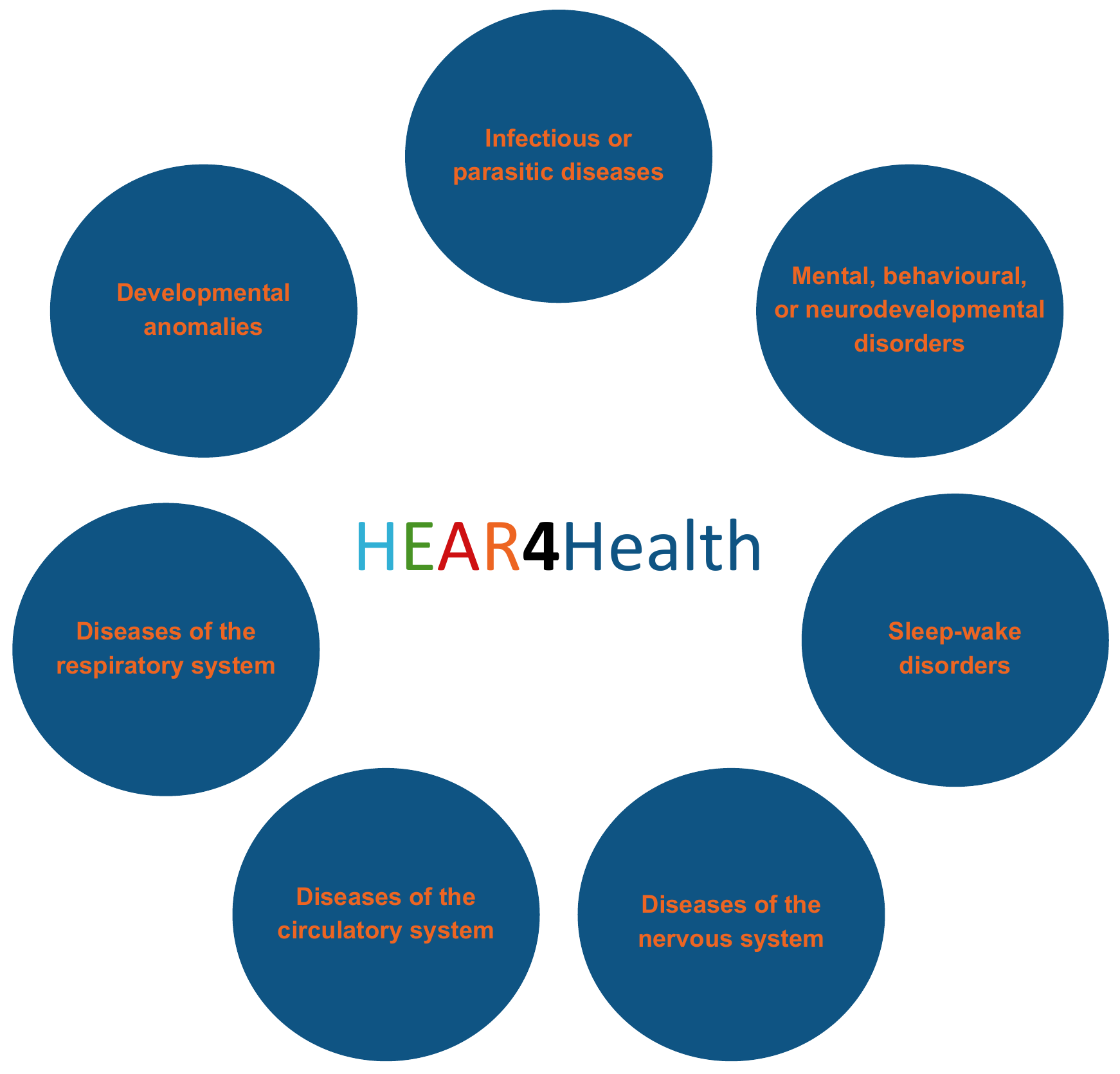}
    \caption{
    ICD-11 categories that are a focal point for computer audition research.
    }
    \label{fig:HEALTH}
\end{figure}

Naturally, any advances in computer audition targeted towards healthcare applications are inextricably tied to the specific medical conditions that lend themselves to modelling via audio; the necessary pre-requisite is that these conditions manifest themselves, at least to some extent, in auditory biomarkers emanating from the patients' body.
Historically, a significantly higher emphasis has been placed on vocalisations compared to other body acoustics such as heart sounds~\citep{Cummins18-SAH}.
Accordingly, this choice has shaped most of the existing approaches and, thus, also becomes the central point of our review.
\Cref{fig:HEALTH} shows the main ICD-11\footnote{https://icd.who.int/} categories on which previous research has focused, ordered clockwise according to their order in the ICD-11 manual.
In the following sections, we proceed to analyse each of those categories, presenting prior computer audition works that have focused on specific diseases, and discussing the impact that our HEAR4Health framework can have on them.

\subsection{Infectious or parasitic diseases}
This broad category covers several communicable diseases, from bacterial, gastrointestinal infections, to sexually transmitted diseases and viral infections, the majority of which do not manifest in auditory biomarkers; the ones that do, however, number several auditory symptoms such as (persistent) coughing or having a sore throat.
The ones predominantly appearing in computer audition literature are: (respiratory) \emph{tuberculosis} (1B10)~\citep{larson2012validation, botha2018detection, ijaz2022towards, zimmer2022making}; \emph{pertussis} (1C12)~\citep{pramono2016cough, imran2020ai4covid}; and \emph{influenza}(1E)~\citep{ward2021flunet}.
Existing works have predominantly focused on detecting and analysing coughs; in particular, the onset of \ac{DL} and the increase in available data have unveiled the potential of detecting coughs and subsequently categorising them as pathological or not.

\subsection{Mental, behavioural, or neurodevelopmental disorders}
The wide variety of symptoms of these diseases, often manifesting as speech and language pathologies, along with their widespread prevalence, have made them prime targets for the computer audition community~\citep{voleti2019review, latif2020speech, miner2020assessing, zhang2022natural}.
Typical disorders are: (developmental) \emph{aphasia} (6A01.20)~\citep{le2014modeling, le2017automatic}; \emph{schizophrenia} (6A20)~\citep{delisi2001speech, tahir2016non, he2021automatic}; \emph{autism} (6A02)~\citep{gernsbacher2016language, rynkiewicz2016investigation, pokorny2017earlier, roche2018early, rudovic2018personalized}; \emph{mood disorders} (6A60; 6A70), of which \emph{depression} is the most commonly researched~\citep{france2000acoustical, cummins2015review, ringeval2019avec}; and \emph{anxiety or fear-related disorders} (6B)~\citep{laukka2008nervous, baird2021evaluation}.
For example, aphasia has been linked to mispronunciation errors and increased effort~\citep{le2014modeling, le2017automatic};
schizophrenia manifests in slower response times in conversations~\citep{tahir2016non}; and blunted affect~\citep{delisi2001speech, tahir2016non, he2021automatic};
depression results in a flatter tone (lower mean F0 values and range) with more pauses~\citep{cummins2015review}, and an increase in jitter and shimmer, indicative of more strained articulation.

\subsection{Sleep-wake disorders}
Research in sleep-wake disorders has been typically targeted to \emph{breathing-disorders} -- mainly apnoeas~\citep{janott2018snoring, korompili2021psg}, while some research has been focused on the detection of the resulting \emph{sleepiness}~\citep{schuller2019interspeech}.
Apnoeas, on the one hand, mostly manifest as very loud snoring, which is caused by a prolonged obstruction of the airways and subsequent `explosive' inspirations.
These signals can be automatically detected and analysed using auditory \ac{ML} systems~\citep{duckitt2006automatic}.
Daytime sleepiness, on the other hand, has been mostly studied as a speech and language disorder; it manifests in lower speaking rates and irregular phonation~\citep{honigautomatic}.

\subsection{Diseases of the nervous system}
This family of diseases has adverse effects on  memory, motor control, and cognitive performance.
Its most widely studied diseases from a speech pathology perspective are Parkinson's (8A00.0)~\citep{Holmes00-VCP, Midi08-VAR}; Alzheimer's (8A20)~\citep{Hoffmann10-TPS, Garcia20-AIS, Luz20-ADR}; multiple sclerosis (8A40)~\citep{Noffs18-WSC}; amyotrophic lateral sclerosis (8B60)~\citep{vieira2022machine}; and cerebral palsy (8D)~\citep{Nordberg14-CPO}.
These manifest primarily in the speech signal, with dysarthria and dysphonia being the most common symptoms. 
For example, studies find that Parkinson's shows up as increased roughness, breathiness and dysphonia, and higher F0 values~\citep{Holmes00-VCP, Midi08-VAR}, Alzheimer's results in more hesitation~\citep{Hoffmann10-TPS}, multiple sclerosis leads to slower and more imprecise articulation, pitch and loudness instability, longer and more frequent pauses~\citep{Noffs18-WSC}, and cerebral palsy shows up as dysarthria, hypernasality, and imprecise articulation of consonants~\citep{Nordberg14-CPO}.

\subsection{Diseases of the circulatory system}
Auscultation has been a mainstay of a medical examination since the invention of the stethoscope by Ren\"e Laennec in 1816, by now a trademark of medical practitioners around the world~\citep{chizner2008cardiac}.
It is particularly useful when listening to the sounds of the heart or the lungs of a patient.
Accordingly, its digital equivalent can be immensely useful in detecting pathologies of the circulatory system, such as arrhythmias or congenital heart diseases.
Analysing those signals has become the topic of multiple PhysioNet challenges~\citep{clifford2016classification} was also featured in the 2018 version of the ComParE series~\citep{schuller2018interspeech}, 
with computer audition systems being developed to detect and classify abnormal events (`murmurs') in phonocardiograms~\citep{singh2007computer, oliveira2021circor}.

\subsection{Diseases of the respiratory system}
These diseases can be broadly taxonomised as being related to the upper or lower respiratory tract.
Prominent examples are bronchitis (CA20)~\citep{imran2020ai4covid}; chronic obstructive pulmonary disease (COPD; CA22)~\citep{Triantafyllopoulos22-DBP, claxton2021identifying}; asthma (CA23)~\citep{kutor2019speech}; pneumonia (CA40)~\citep{kosasih2014wavelet}; and COVID-19 (RA01; designated under `codes for special purposes' due to the pandemic emergency)~\citep{deshpande2022ai, han2022sounds}.
By nature of their symptomatology, these diseases are prototypical examples of ones that manifest in auditory biomarkers.
Thus, different signals have been used to detect their presence, such as speech~\citep{Triantafyllopoulos22-COVYT}, breathing and coughing~\citep{sharma2020coswara, brown2020exploring}, or sustained vowels~\citep{bartl2021voice}.

\subsection{Developmental anomalies}
Developmental disorders, such as the Angelman syndrome (LD90.0), Rett syndrome (LD90.4), and fragile X syndrome (LD55) manifest in divergent vocalisation and speech development patterns from an early age~\citep{roche2018early, grieco2018quantitative, bartl2022vocalisation, pokorny2022automatic}.
Infants with specific developmental disorders produce abnormal cooing sounds and less person-directed vocalisations, and their vocalisations are found to be of lower complexity as compared to typically developing infants.
From a signal perspective, these anomalies manifest in speech, first in pre-linguistic sounds and later on in linguistic vocalisations of young children.
As the emphasis is on children and young adults, they present an additional challenge to data collection, on the one hand due to ethical and privacy reasons, and on the other due to a potentially reduced compliance of children with recording requirements.


\section{HEAR4Health: A blueprint for future auditory digital health}

Unifying the four pillars results in a working digital health system which we name \emph{HEAR}.
Our system can be used to supplement the decision-making of practitioners across a wide facet of diseases.
In general, we anticipate two distinct functioning modes for it.
On the one hand, it can be used as a general-purpose \emph{screening} tool to monitor healthy individuals and provide early warning signs of a potential disease.
This hearing with `all ears open' mode takes a holistic approach, and emphasises a wide coverage of symptoms and diseases, thus functioning as an early alarm system that triggers a follow-up investigation.
Following that, it can be utilised to \emph{monitor} the state of patients after they have been diagnosed with a disease, or for measuring the effect of an intervention.
This second, more constrained setting necessitates a `human-in-the-loop' paradigm, where the doctor isolates a narrower set of biomarkers for the system to monitor --now with more focus and prior information about the patient's state-- which is then reported back for each new follow-up.

In either case, there are stringent requirements for reliability and explainability that can only be satisfied with the use of prior knowledge, attention to the individual, and an adherence to ethical principles.
Ultimately, it is user trust that is the deciding factor behind the adoption of a transformative technology.
The use of computer audition in healthcare applications is currently in its nascent stages, with a vast potential for improvement.
Our blueprint, \emph{HEAR4Health}, incorporates the necessary design principles and pragmatic considerations that need to be accounted for by the next wave of research advances to turn audition into a cornerstone of future, digitised healthcare systems.

\section{\refname}
\printbibliography[heading=none]

\end{document}